\documentclass[%
 reprint,
 amsmath,amssymb,
aps,
pra,
]{revtex4-2}

\usepackage{graphicx}
\usepackage{dcolumn}
\usepackage{bm}
\usepackage{color}

\begin{document}

\hyphenpenalty=2000
\preprint{APS/123-QED}

\title{Nondipole effects on the double-slit interference in molecular ionization by xuv pulses}

\author{Kunlong Liu}
\email{liukunlong@hust.edu.cn}
\affiliation{School of Physics, Huazhong University of Science and Technology, Wuhan 430074, China}

\author{Yibo Hu}
\affiliation{School of Physics, Huazhong University of Science and Technology, Wuhan 430074, China}

\author{Qingbin Zhang}%
\affiliation{School of Physics, Huazhong University of Science and Technology, Wuhan 430074, China}

\author{Peixiang Lu}%
\affiliation{School of Physics, Huazhong University of Science and Technology, Wuhan 430074, China}

\date{\today}

\begin{abstract}
The double-slit interference in single-photon ionization of the diatomic molecular ion $\mathrm{H}_2^+$ is theoretically studied beyond the dipole approximation. Via simulating and comparing the interactions of the prealigned $\mathrm{H}_2^+$ and the hydrogen atom with the xuv pulses propagating in different directions, we illustrate
two kinds of effects that are encoded in the interference patterns of the photoelectrons from $\mathrm{H}_2^+$: (i) the photon-momentum transfer and (ii) the finite speed of light. While both effects could modify the maxima of the interference fringes, we show that the former one hardly affects the interference \textit{minima}. 
Our results and analysis show that the interference minima rule out the influences of the photon-momentum transfer and, potentially, the multielectron effect, thus performing a better role in decoding the zeptosecond time delay for the pulse hitting one and the other atomic centers of the molecule.
\end{abstract}

\maketitle


\section{Introduction}
The development of few-cycle infrared laser pulses, attosecond pulses, and free-electron lasers in the past two decades has enabled scientists to detect and control the ultrafast electronic and nuclear dynamics on the attosecond time scale (see \cite{atto} and the references therein). 
While our understanding of the dynamic mechanisms of single atoms and molecules driven by strong fields is getting deeper and deeper, researchers have already been seeking new challenges in strong-field physics. On one hand, for instance, efforts have been made in studying the high harmonic generation in chiral molecules \cite{chiral}, liquids \cite{liq}, and solids \cite{solid1} as well as the photoemission process in condensed matter \cite{cond1} and nano-particles \cite{nanop}, bringing us unprecedented insights into the strong-field phenomena in complex systems. On the other hand, scientists have been exploring the ever-faster responses of atoms and molecules below one attosecond \cite{zept}. 

Recently, the zeptosecond ($10^{-21}$ s) time delay between the ionization bursts from two centers of a diatomic molecule was measured \cite{zept2}. In the experiment, the 800-eV xuv laser pulse was applied to trigger the single-photon ionization of the hydrogen molecule. Due to the finite speed of light, it takes time for the pulse to hit one and the other nuclei of the molecule, leading to a phase difference between the ionizing wave packets from two centers and eventually resulting in the distorted interference patterns in the photoelectron momentum distributions (PMDs). In principle, the information of the time delay could be extracted from the tilted angle of the double-slit interference maximum (theoretical details will be discussed in the discussion section). Yet, a noticeable deviation was found between the measurement and the theoretical prediction for the time delay \cite{zept2}. The deviation, as well as the cutting-edge experimental research itself, has immediately attracted broad interest and discussion \cite{z1,z2,z3}. 

The nondipole effect in molecular strong-field ionization has been previously studied in theory \cite{ndm1,ndm2,ndm3}, mainly focusing on the photoelectron momentum offset in the light propagation direction. Such momentum-transfer effect takes place in atomic ionization as well, and it will be referred to as the single-atom nondipole effect in the following discussion.
In this paper, we revisit the double-slit interference in the single-photon ionization of the diatomic molecules \cite{fano} by numerically solving the three-dimensional time-dependent Schr\"odinger equations beyond the dipole approximation. 
We focus on the two-center-interference nondipole effect induced by the time delay for the pulse hitting one and the other nuclei.
By comparing the PMDs for the hydrogen atom and the hydrogen molecular ion, we illustrate how the photon recoil affects the interference patterns when the pulse is travelling at different directions. In particular, our results and analysis show that the interference minima are associated with the time delay for the pulse travelling from one to the other nuclei, but they would be hardly affected by the single-atom nondipole effect. We demonstrate that applying the two neighboring minima of the zeroth-order interference maxima is more accurate for extracting the zeptosecond time delay between the ionization bursts from two atomic centers of the molecule.

\section{Numerical methods}

\begin{figure}[b]
\begin{center}
\includegraphics[width=6.0 cm]{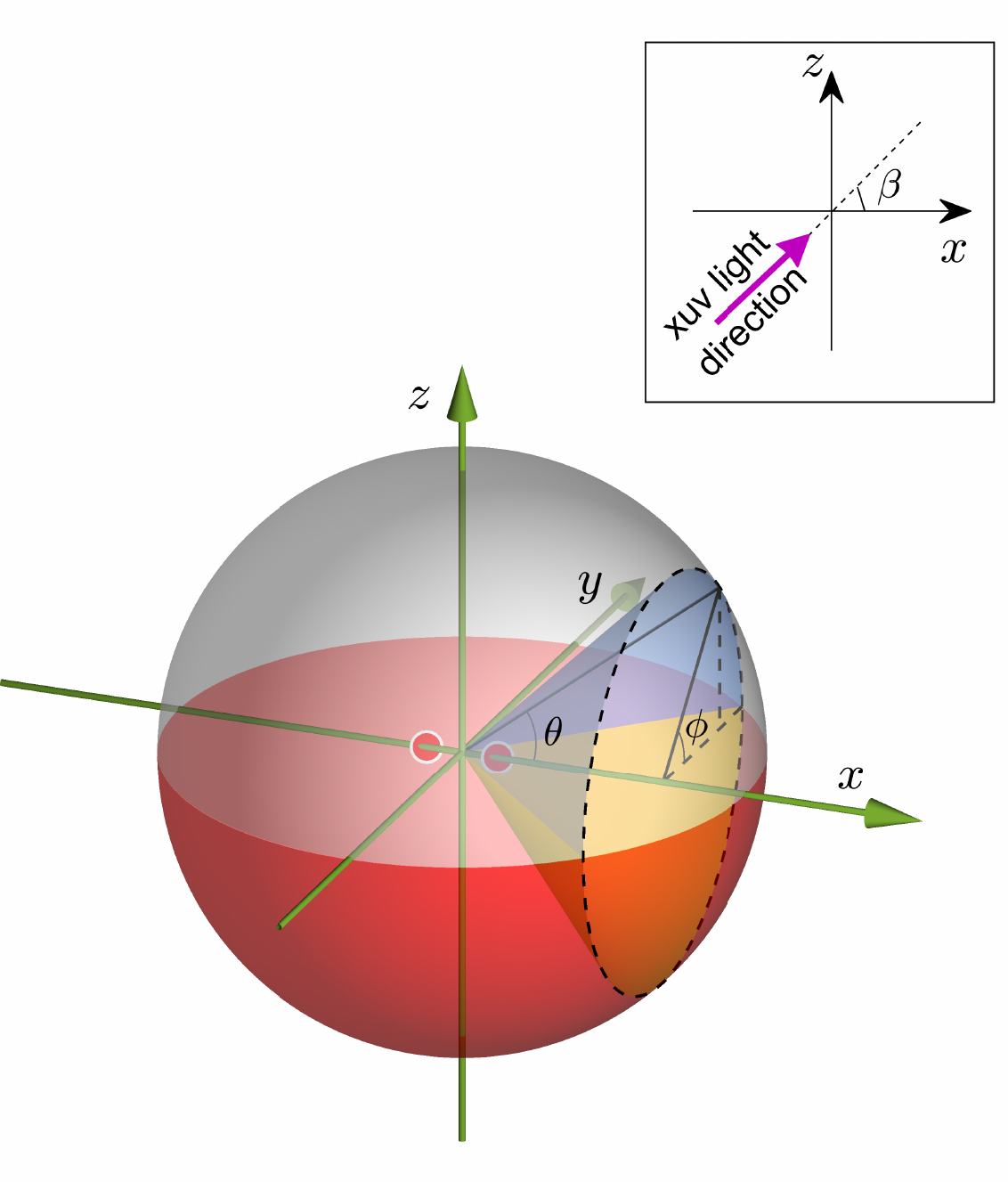}
\caption{\label{fig:xyz} Illustration of the coordinates in our simulation. The molecule is prealigned along the $x$ axis. The xuv laser pulse is linearly polarized in the $y$ direction and propagates in the direction given by the angle $\beta$. The angular distributions of the photoelectrons are calculated as a function of $\theta$ ($0 \leq \theta \leq \pi$) and $\phi$ ($0 \leq \phi < 2\pi$) indicated in the figure.}
\end{center}
\end{figure}

We solve numerically the nondipole three-dimensional (3D) time-dependent Schr\"odinger equations (TDSEs) in Cartesian coordinate system for the H$_2^+$ prealigned along the $x$ axis as well as for the hydrogen atom.
The xuv laser pulse is linearly polarized in the $y$ direction and propagates in the direction of $\mathbf{e}_{k}=\cos \beta \mathbf{e}_{x}+ \sin \beta \mathbf{e}_{z}$, where $\beta$ is the angle between the propagation direction and the $x$ axis, as shown in the sub-diagram in Fig.~\ref{fig:xyz}.
The TDSE is given by (in atomic units) \cite{JMOtdse}
\begin{eqnarray}
\label{eq:TDSE}
i\frac{\partial}{\partial t}\Psi(\mathbf{r},t)
= \left\{
\frac{1}{2}\left[-i\nabla+\mathbf{A}(\eta)\right]^2+V_{0}(\mathbf{r})
\right\}\Psi(\mathbf{r},t),\ \
\end{eqnarray}
where $V_{0}(\mathbf{r})$ is the Coulomb potential of the chosen system.
The vector potential of the xuv laser pulse is given by
\begin{eqnarray}
\mathbf{A}(\eta) = A_y(\eta)\mathbf{e}_{y}=
\frac{\mathcal{E}}{\omega}\sin^4\left(\frac{\pi \eta}{\omega NT}\right)
\cos( \eta)\mathbf{e}_{y},
\end{eqnarray}
for $0 \leq \eta \leq \omega NT$ and otherwise ${A}(\eta)=0$, with $\eta = \omega [t- (x\cos\beta+z\sin\beta)/c]$, where $c$ is the speed of light in \textit{vacuum}.
Here,
$\omega$, $T=2\pi/\omega$, $\mathcal{E}$, and $N$ indicate the laser frequency, the optical cycle, the electric field amplitude, and the number of the optical cycles of the full pulse, respectively.
In the present simulations, the laser parameters are chosen as $\omega=29.40$ a.u. (corresponding to the photon energy of $800$ eV), $\mathcal{E}=3$ a.u., and $N=50$.

The nondipole TDSE is numerically solved using the split-operator spectral method \cite{fft} with modifications, in order to deal with the space-dependent vector potential. The details of the procedure are following.
The analytic solution of Eq.~(\ref{eq:TDSE}) is given by a Dyson's time ordering operator $\mathbf{P}$ as (in the Coulomb gauge)
\begin{widetext}
\begin{eqnarray}
\label{eq:Dyson}
\Psi(\mathbf{r},t+\delta t) &=&\mathbf{P}
\exp \left\{
-i \int^{t+\delta t}_t d\tau
\left[  \frac{1}{2}\left(-i\mathbf{\nabla}+\mathbf{A}(\eta)\right)^2+V_{0}(\mathbf{r}) \right]
\right\}
\Psi(\mathbf{r},t) \\ \nonumber
&=&\mathbf{P}
\exp \left\{
-i \int^{t+\delta t}_t d\tau
\left[  
\frac{1}{2}(-i\mathbf{\nabla})^2 + \mathbf{A}(\eta)\cdot (-i\mathbf{\nabla})+\frac{1}{2} \mathbf{A}^2(\eta)+V_{0}(\mathbf{r})
\right]
\right\}
\Psi(\mathbf{r},t) \\ \nonumber
&=&\mathbf{P}
\exp \left\{
-i \int^{t+\delta t}_t d\tau
\left[  
\frac{1}{2}\hat{\mathbf{p}}^2  + 
{A}_y(x,z,\tau) \hat{p}_y+\frac{1}{2}
\mathbf{A}^2(\mathbf{r},\tau)+V_{0}(\mathbf{r}) 
\right]
\right\}
\Psi(\mathbf{r},t),
\end{eqnarray}
\end{widetext}
with the momentum operator $\hat{\mathbf{p}}=-i\mathbf{\nabla}=(\hat{{p}}_x,\hat{{p}}_y,\hat{{p}}_z)$.
The approximate evaluation of the exponential with time ordering in Eq.~(\ref{eq:Dyson}) can be written as \cite{split}
\begin{eqnarray} 
\label{eq:split}
\Psi(\mathbf{r},t+\delta t) &\approx &
\exp \left\{-i\frac{\delta t}{2}
\left[ \frac{1}{2}
\mathbf{A}^2\left(\mathbf{r},t'\right)  +V_{0}(\mathbf{r}) \right]
\right\} \nonumber \\ \nonumber
&& \times 
\exp \left\{-i\frac{\delta t}{2}
\left[ 
{A}_y(x,z,t')\hat{p}_y
\right]
\right\} \\ \nonumber
&& \times 
\exp \left\{-i \delta t 
\frac{\hat{\mathbf{p}}^2 }{2}
\right\} \\ \nonumber
&& \times 
\exp \left\{-i\frac{\delta t}{2}
\left[ 
{A}_y(x,z,t')\hat{p}_y 
\right]
\right\} \\  \nonumber
&& \times
\exp \left\{-i\frac{\delta t}{2}
\left[ \frac{1}{2}
\mathbf{A}^2\left(\mathbf{r},t'\right)  +V_{0}(\mathbf{r}) \right]
\right\} \\
&& \times \Psi(\mathbf{r},t)   
\end{eqnarray}
with $t'=t+\delta t/2$. The first exponential operation on the wave function can be calculated directly and one can obtain a temporary wave function. Then, the $y$ dimension of the temporary wave function is Fourier transformed to the momentum space as $\Psi(x,y,z)\rightarrow \Psi(x,p_y,z) $, so that the second operation can be calculated by direct multiplication. In the next step, the other two dimensions of the wave function are transformed to the momentum space as $\Psi(x,p_y,z)\rightarrow \Psi(p_x,p_y,p_z) $, for the calculation of the third operation. Finally, the rest of the operations can be analogously evaluated via transforming the temporary wave function back to the position space step by step. By repeating the processes above, one can obtain the evolution of the wave function numerically.

\begin{figure*}[t]
\begin{center}
\includegraphics[width=18cm]{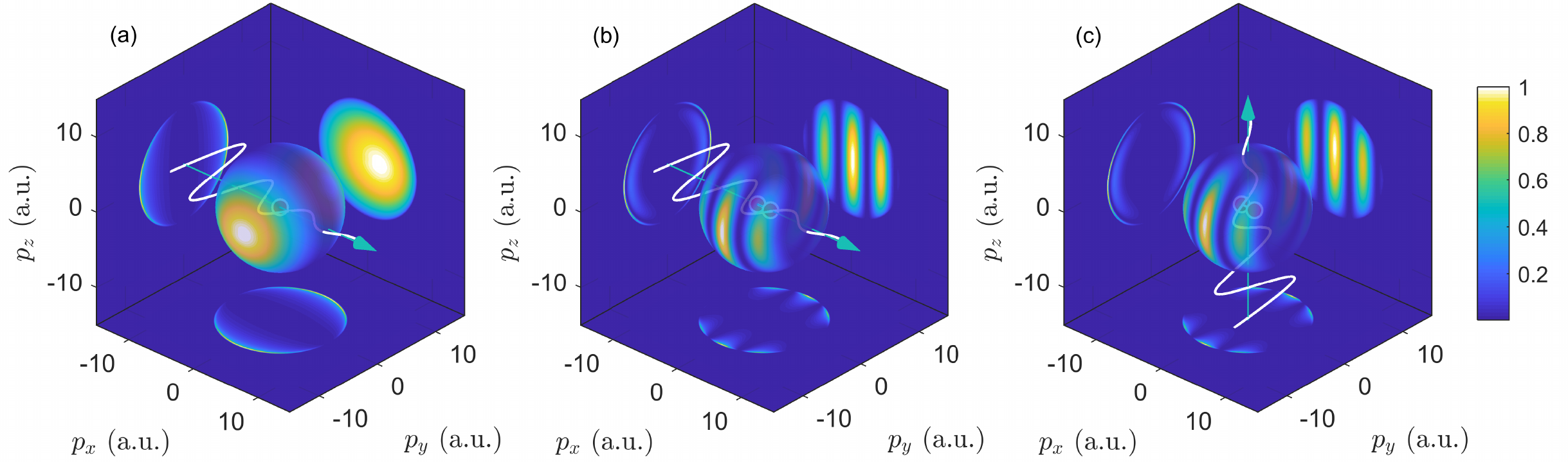}
\caption{\label{fig:PMD4d} The momentum and angular distributions of the photoelectrons for the interactions of the atom (a) and the molecule ($R=1.4$ a.u.) [(b) and (c)] with the xuv pulses. The oscillating curves illustrate the electric components of the laser pulses and the arrows indicate the laser propagation directions. The distributions are normalized to the cooresponding maximal values and the color scale is linear.}
\end{center}
\end{figure*}

The initial stationary wave functions are obtained by the imaginary-time propagation method.
The real-time propagation includes two parts: the interaction and the free propagation afterwards.
For the interaction part, the evolution starts when the xuv pulse enters the box of the 3D grid and ends when the tail of the pulse lefts the box. We have chosen a 3D grid large enough to contain the majority of the ionizing wave packets until the interaction process ends. This is guaranteed and has been verified in our calculations. After the interaction, we continue the evolution of the wave function without external fields and apply the absorbing potential to split the outgoing wave packets. In this case, one can solve the TDSE and obtain the photoelectron momentum distributions in the conventional way \cite{eiLiu,pctdse}.
Regarding the simulation parameters,
there are $2496\times2496\times520$ grid points of the box. The spacing steps are $\Delta x=\Delta y=1/6$ a.u.~for $R=2$ and $10$ a.u., $\Delta x=\Delta y=0.14$ a.u.~for $R=1.4$ a.u., and $\Delta z=0.25$ a.u. The spacing steps are chosen differently for the given internuclear distances $R$,
so that the nuclei are placed in the middle of two grid points in the $x$ axis to avoid the singularity of the Coulomb potential \cite{pctdse}.
The time step for the evolution of the wave function is chosen as $\delta t=0.005$ a.u.

\section{Results and discussion}
We have calculated the 3D photoelectron momentum distributions $Y_\mathrm{3D}$ for the hydrogen atom and H$_2^+$ interacting with the xuv pulses. We show in Fig.~\ref{fig:PMD4d} the results of three chosen cases: (a) for the atom and $\beta=0$, (b) for H$_2^+$ and $\beta=0$, and (c) for H$_2^+$ and $\beta=90^\circ$, with $R=1.4$ a.u. 
The colored spheres in the figures indicate the angular distributions 
\begin{eqnarray}
Y_\text{ang}(\theta,\phi)=\int Y_\mathrm{3D}(p,\theta,\phi) p^2dp 
\end{eqnarray}
of the photoelectrons (see Fig.~1 for the definition of $\theta$ and $\phi$). 
We also show in the figures the corresponding 2D distributions, $Y_{xy}(p_x,p_y)$ on the bottom, $Y_{yz}(p_y,p_z)$ on the left-hand site, and $Y_{xz}(p_x,p_z)$ on the right-hand site, which are integrated over the third dimensions of $Y_\mathrm{3D}(p_x,p_y,p_z)$, respectively. As shown in Fig.~\ref{fig:PMD4d}(a), the peak of the distribution $Y_{xz}$ drifted towards the light propagation direction, due to the momentum transfer from the photon to the photoelectron. For the cases of H$_2^+$ shown in Figs.~\ref{fig:PMD4d}(b) and \ref{fig:PMD4d}(c), the double-slit interference patterns appear. 
The interference fringes shown by $Y_{xz}$ are perpendicular to the molecular axis for both cases, but the distributions of the fringes seem different.

To quantitatively compare the distributions of the interference patterns, we calculated the photoelectron angular distributions (PADs) given by
\begin{eqnarray}
&&Y_\theta^+(\theta)=\int_0^\pi Y_\text{ang}(\theta,\phi) d\phi \\
&&Y_\theta^-(\theta)=\int_\pi^{2\pi} Y_\text{ang}(\theta,\phi) d\phi
\end{eqnarray}
for the atom and H$_2^+$ ($R=1.4$ a.u.) in the cases of $\beta=0$, $45^\circ$, and $90^\circ$, respectively. The results after normalization to the maximum of $Y_\theta^+$ are shown in Figs.~3(a)--3(c).
We can see that PADs for the interference patterns are enveloped by those PADs for the hydrogen atom. 
It indicates that, when the xuv pulse hits each center of the molecule, the ionizing wave packet burst is approximately equivalent to that from a single atom and two wave packets coherently interfere with each other in the continuum. 
Thus,
the distorted PAD from each atomic center will be encoded in the interference patterns, modifying the locations and yields of the fringe maxima. This can be seen in both Fig.~2 and Fig.~3.

\begin{figure*}[t]
\begin{center}
\includegraphics[width=18cm]{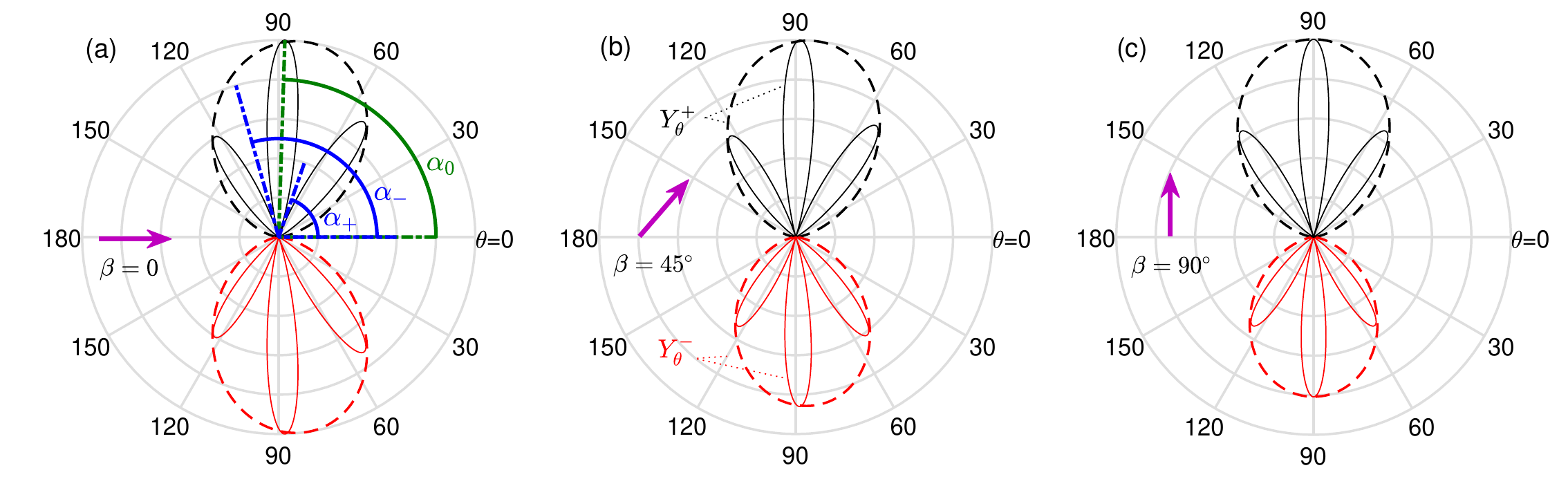}
\caption{\label{fig:PAD} The photoelectron angular distributions given by $Y_\theta^\pm$ for the atom (dashed curves) and the molecule
($R=1.4$ a.u.)
(solid curves) interacting with the pulses propagating in three directions (indicated by the arrows). 
$\alpha_0$ and $\alpha_\pm$ indicate the angles where the interference maximum and its two neighboring minima are observed, respectively.}
\end{center}
\end{figure*}

When the pulse travels at $\beta=90^\circ$ (i.e., perpendicular to the molecular axis), the photon-momentum transfer leads to more photoelectron yield for $p_z>0$ than that for $p_z<0$. The global maxima of $Y_\theta^+$ and $Y_\theta^-$ are no longer equal, although they both appear at $\theta=90^\circ$ for the atom and the molecule, as shown in Fig.~\ref{fig:PAD}(c). In this particular case, the pulse interacts with two atomic centers at the same time.
The situation becomes complicated when the pulse no longer travels perpendicular to the molecular axis. 
As shown in Figs.~\ref{fig:PAD}(a) and \ref{fig:PAD}(b), the PADs for the atom are tilted towards the light direction and they peak at the angle (referred to as $\theta_A$) far from $\theta=90^\circ$. For the molecule, the interference maximum is also tilted but peaks at the angle (referred to as $\alpha_0$) slightly smaller than $90^\circ$. At a first glance, the tilted interference maximum is affected by the photon-momentum transfer during the interaction. This is true, but it was shown that the finite speed of light plays a more important role in modifying the interference patterns \cite{zept2}. 
In details, beyond the dipole approximation, there is a time delay for the pulse interacting with one and the other atomic centers of the molecule due to the finite speed of light, resulting in a phase difference between the ionizing wave packets generated from two centers. Such phase difference eventually modifies the double-slit interference pattern of the photoelectrons.

It was proposed that by finding the interference maxima, one can obtain the interacting time delay that is as small as hundreds of zeptoseconds \cite{zept2}. However, as illustrated in Fig.~\ref{fig:PAD} and also shown by previous works \cite{z2}, the photon-momentum transfer has impact on the interference maxima.
Moreover, in multielectron systems, the multielectron dynamics would potentially affect the interference patterns, as discussed in \cite{z1}.
These factors might limit the accuracy of extracting the time delay from the interference maximum.
In the following, we will show that adopting the interference minima instead of the maxima will rule out the side effects and provides us with a better way to decode the interacting time delay.

We start with the wave function of the photoelectron ionizing from the atom, which is written as
\begin{eqnarray}
\label{eq:psia}
\Psi_\text{A}(\mathbf{p})=A(\mathbf{p})\exp[i(\mathbf{p}\cdot \mathbf{r}-E_k t)]
\end{eqnarray}
where $A(\mathbf{p})$ describes the amplitude of the yield distribution and $E_k=\mathbf{p}^2/2$ indicates the final kinetic energy of the photoelectron. In Eq.~(\ref{eq:psia}), we have made two assumptions: (A1) the electronic wave packet originates from the origin and (A2) the Coulomb effect on the spreading wave packet is neglectable.  
By further assuming that the ionizing wave packet generated from each center of the molecule is equivalent to that from a single atom (referred to as the assumption A3), we can write the wave function for the photoelectron emitted from the left and right atomic centers, respectively, as
\begin{eqnarray}
\label{eq:psiL}
\Psi_\text{L}(\mathbf{p})&=&A(\mathbf{p})\exp\{i[\mathbf{p}\cdot \mathbf{r}_1-(E_k t+ \omega t_d)]\}, \\
\Psi_\text{R}(\mathbf{p})&=&A(\mathbf{p})\exp[i(\mathbf{p}\cdot \mathbf{r}_2-E_k t)],
\end{eqnarray}
with $\mathbf{r}_{1,2}=\mathbf{r} \pm \mathbf{R}/2$ and $\mathbf{R}=(R,0,0)$. 
In Eq.~(\ref{eq:psiL}), $t_d$ indicates the interacting time delay.
The term $ \omega t_d$ is the extra phase difference between the wave packets generated from the left and right atomic centers due to the time delay. 
This phase difference can be understood as following. We assume that the xuv pulse interacts with the atom on the left side earlier. Then, by absorbing the photon energy, the total energy of the wave packet about to be ionized from the left core is lifted by $\omega$, while that on the right side remains unchanged. When a duration of $t_d$ passes, an initial phase difference of 
$\omega t_d$ between two sides arises. After the interaction, the ionizing wave packets from both sides experience the same process: carrying the same energy, escaping from the parent core, and ending up with the same kinetic energy, which will accumulate the same phase for both wave packets. Thus, the overall phase difference (regarding the energy part) between $\Psi_L$ and $\Psi_R$ is $\omega t_d$. 

\begin{figure}[t]
\begin{center}
\includegraphics[width=8.6 cm]{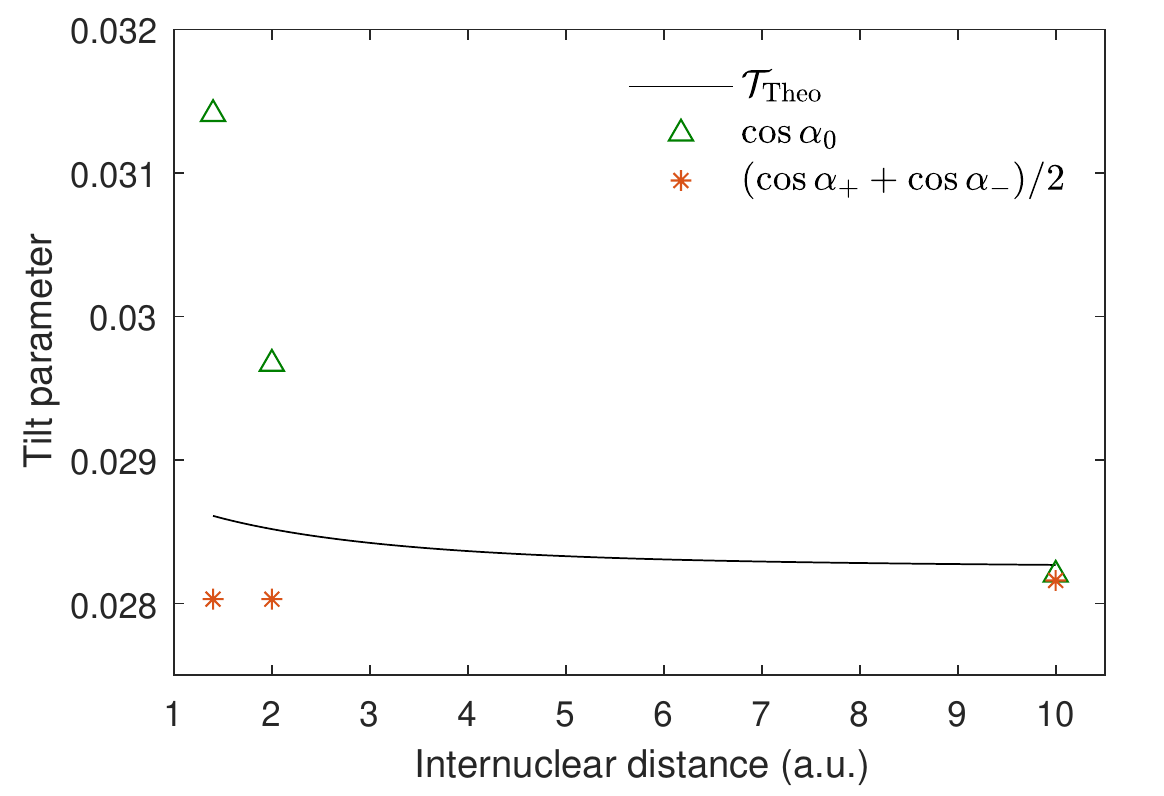}
\caption{\label{fig:alpha} The tilt parameter as a function of the molecular internuclear distance in the case of $\beta=0$. The locations of the maxima and the minima are obtained with the spline interpolation of the photoelectron angular distributions.}
\end{center}
\end{figure}

Since we have $E_k=\omega-I_p$ in single-photon ionization, with $I_p$ being the ionization potential, we define $E_0=E_k+I_p \equiv \omega$ as the initial kinetic energy of the photoelectron. 
Then, the wave function of the photoelectron from the diatomic molecules is written as
\begin{eqnarray}
\label{eq:psim}
\Psi_\text{M}(\mathbf{p}) 
&\approx&  \Psi_\text{L}(\mathbf{p})+\Psi_\text{R}(\mathbf{p})  \nonumber \\
 &=& A(\mathbf{p})\exp[i(\mathbf{p}\cdot \mathbf{r}-E_k t)]    \nonumber \\
&& \times \exp\left(-i \mathbf{p} \cdot \frac{\mathbf{R}}{2}\right)   \nonumber \\
&& \times \{  1+ \exp [ i(\mathbf{p}\cdot \mathbf{R}-E_0 t_d) ]  \}.
\end{eqnarray}
On the righthand site of Eq.~(\ref{eq:psim}), the first line equals to $\Psi_\text{A}(\mathbf{p})$, i.e.~the photoelectron wave function from a single atom. The second one simply introduces an extra global phase to the wave function. The last one is the interference term which modulates the PADs. 
In principle, the interacting time delay can be decoded from the interference pattern as $t_d$ appears in the interference term.
However, it could be tricky for measurements. Let us look into the conditions for the extrema of the interference.
The interference term leads to its maxima when the following condition is satisfied:
\begin{eqnarray}
\label{eq:max}
\exp [ i(\mathbf{p}\cdot \mathbf{R}-E_0 t_d) ] =1  . 
\end{eqnarray}
For the given photoelectron momentum $p=\sqrt{2(\omega-I_p)}$ with $I_p$ being the ionization potential of the system, 
the interference maxima can be found at the angles $\theta_n$ satisfying
\begin{eqnarray}
\label{eq:thetan}
pR\cos\theta_n -E_0 t_d= n\cdot(2\pi).
\end{eqnarray}
In particular, for the zeroth-order interference maximum (i.e., $n=0$), we define the tilt parameter as
\begin{eqnarray}
\label{eq:theta0}
\mathcal{T}_\mathrm{Theo}\equiv \cos\theta_0=\frac{E_0 t_d}{pR} = \frac{E_0}{pc} \cos\beta
\end{eqnarray}
by assuming $t_d=(R\cos\beta)/c$.
Note that the conditions shown above only guarantee the maxima of the interference term in Eq.~(\ref{eq:psim}). The global maximum of the PAD relies on the envelope $A(\mathbf{p})$ as well. In general cases, $A(\mathbf{p})$ peaks beyond $\theta_0$, unless the pulse travels perpendicular to the molecular axis, as already shown in Fig.~\ref{fig:PAD}. 
Therefore, the angle $\alpha_0$ extracted from the maximum of the observed interference fringe is expected to differ from $\theta_0$. 
To verify our expectation, 
we compare the values of 
$\mathcal{T}_\mathrm{Theo}$ and $\cos\alpha_0$ (extracted from the simulated PADs) at three internuclear distances in the case of $\beta=0$. Note that for the calculation of $\mathcal{T}_\mathrm{Theo}$, we have assumed $E_0=\omega$, i.e., the initial kinetic energy of the photoelectron equals to the photon energy.
The results are shown in Fig.~\ref{fig:alpha}.
It is noticeable that $\cos\alpha_0$ tends to deviate from $\mathcal{T}_\mathrm{Theo}$ more significantly as the internuclear distance becomes smaller. 
This can be intuitively understood from Fig.~5, where we show the PADs obtained from numerical solutions of the TDSEs for $R=1.4$ and $10$ a.u. and $\beta=0$. For larger $R$, the frequency of the modulation of the interference fringes becomes higher. In this case, the interference fringes are so narrow that the tilted envelope hardly modifies the interference maxima. 
For $R=1.4$ a.u., however, the broad distribution of the interference fringes would be affected more significantly, leading to the deviation of $\cos\alpha_0$ from $\mathcal{T}_\mathrm{Theo}$.

\begin{figure}[t]
\begin{center}
\includegraphics[width=8.6 cm]{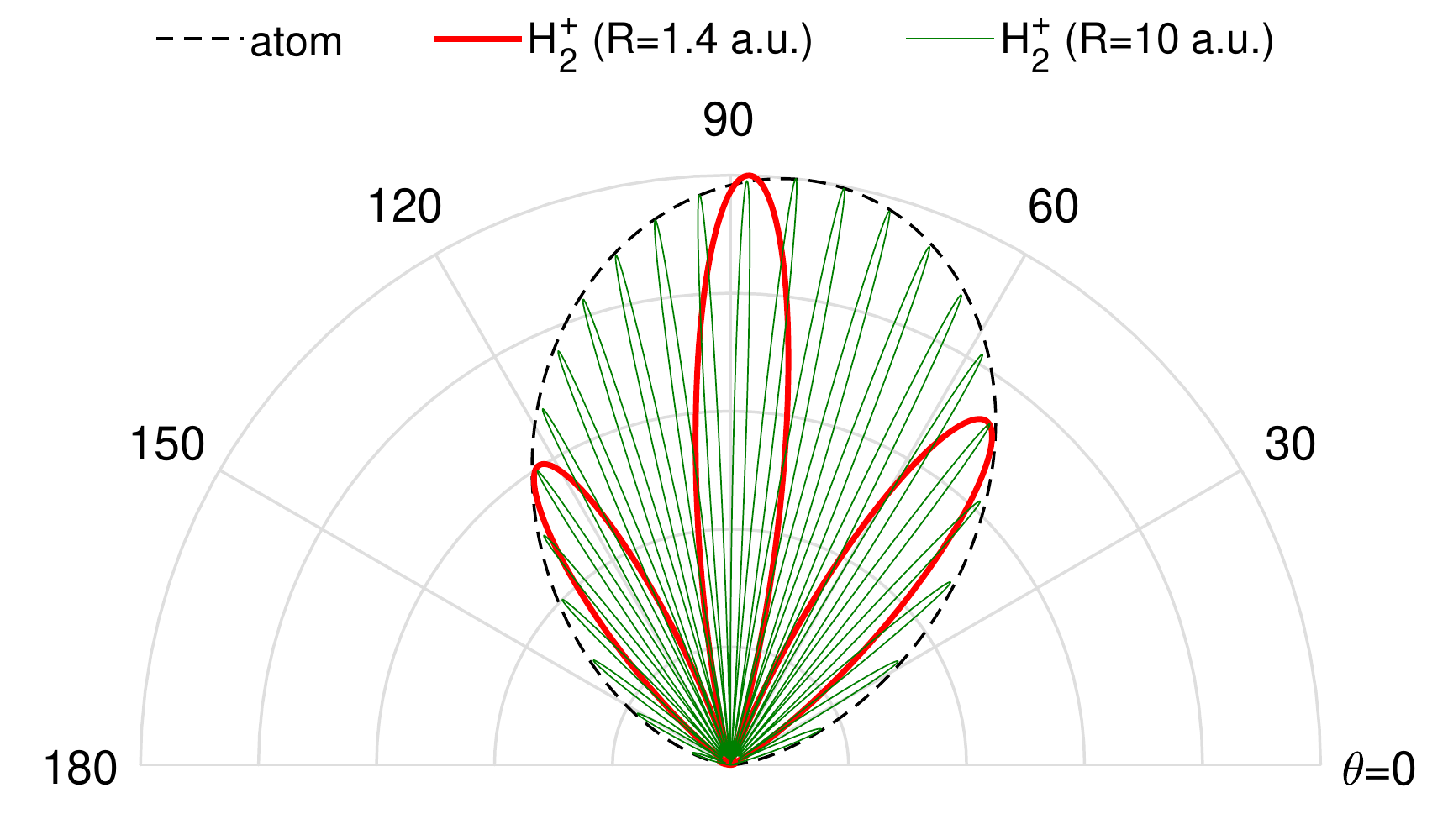}
\caption{\label{fig:padrr} The photoelectron angular distributions for the atom and the molecules of $R=1.4$ and $10$ a.u. in the case of $\beta=0$.}
\end{center}
\end{figure}

\begin{figure*}[t]
\begin{center}
\includegraphics[width=17.5cm]{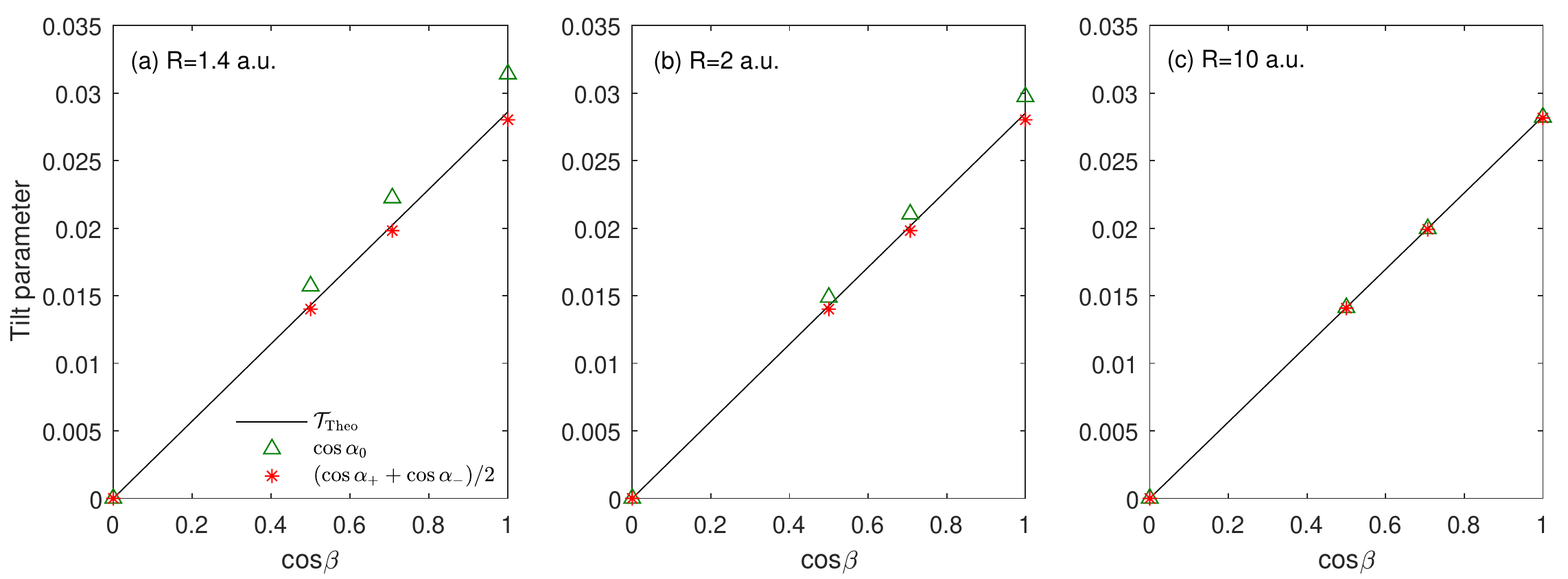}
\caption{\label{fig:3R} The tilt parameter as a function of $\cos\beta$ at three internuclear distances.}
\end{center}
\end{figure*}

To rule out the effect of the envelope $A(\mathbf{p})$, we turn to the interference minima. According to Eq.~(\ref{eq:psim}), the condition for the interference minima is given by 
\begin{eqnarray}
\label{eq:min}
\exp [ i(\mathbf{p}\cdot \mathbf{R}-E_0 t_d) ] =-1  . 
\end{eqnarray}
The two neighboring minima of the zeroth-order maxima are located at the angles $\theta_\pm$ given by
\begin{eqnarray}
\label{eq:thetani}
pR\cos\theta_\pm -E_0 t_d= \pm\pi.
\end{eqnarray}
Then, we can obtain the relation between $\theta_\pm$ and $\theta_0$ as following:
\begin{eqnarray}
\label{eq:maxmin}
\frac{1}{2}(\cos\theta_+ +\cos\theta_-)=\frac{E_0 t_d}{pR}= \frac{E_0}{pc} \cos\beta = \cos\theta_0\equiv \mathcal{T}_\text{Theo}.
\end{eqnarray}
Theoretically, the interference term in Eq.~(\ref{eq:psim}) equals to zero at $\theta_\pm$, 
so the effect of the envelope $A(\mathbf{p})$ on the interference minima should be vanishing. 
For demonstration, we extract the angles $\alpha_\pm$ for the neighboring minima of the zeroth-order maxima from the PADs and show the tilt parameter given by $(\cos\alpha_+ +\cos\alpha_-)/2$ 
in Fig.~\ref{fig:alpha}. One can see that 
the theoretical prediction is generally in agreement with the observation given by $(\cos\alpha_+ +\cos\alpha_-)/2$. In particular,
the deviation for small internuclear distances is reduced remarkably.

In Figs.~\ref{fig:alpha}, we notice that $ \mathcal{T}_\mathrm{Theo}$ is slightly higher than the tilt parameter given by $(\cos\alpha_+ +\cos\alpha_-)/2$. One of the possible reasons is that we assumed a constant initial kinetic energy equal to the photon energy in Eqs.~(\ref{eq:psiL}) and (\ref{eq:maxmin}). In fact, the photoelectron experiences a decelerating process even in the duration of $t_d$, and the deceleration would become more pronounced under the stronger Coulomb attraction. As a result, the theoretical $\mathcal{T}_\mathrm{Theo}$ has been overestimated by adopting $E_0=\omega$ in Eq.~(\ref{eq:theta0}), especially for smaller internuclear distances. 
Besides, the assumptions A1--A3 mentioned in the previous discussion are also the potential reasons leading to the deviation between the theoretical prediction and the numerical calculations. Nevertheless, the assumptions A1--A3 are fairly justified as the deviation is less than $1.6\%$.

Furthermore, the theoretical model given by Eq.~(\ref{eq:theta0}) indicates that the tilt parameter is linearly proportional to $\cos\beta$. 
In Fig.~\ref{fig:3R}, we show the dependence of the tilt parameter on the direction of the light propagation. We can see that the tilt parameters obtained from the interference minima of the PADs agree with the theoretical results very well for all three internuclear distances.

So far, the present work based on the single-active-electron system does not reproduce the results measured in the experiment \cite{zept2}, even if we look at the tilt parameters given by the interference maxima (see Fig.~\ref{fig:3R}). 
On the bright side, however, we have shown that the interference minima of the observable PADs perform a better role in extracting the information regarding the interacting time delay, as they mathematically rule out the influence from the photon-momentum transfer. 
The remaining question is whether the multielectron effect could be ruled out as well at the interference minima. 
To answer this question, let us revisit the theoretical model given by Eq.~(\ref{eq:psim}). 
On one hand, 
the photon-momentum transfer relies on the Coulomb interaction of the photoelectron with the ion, as shown by previous studies \cite{ndm1}. 
If we further consider an additional active bound electron remaining in the ion, then 
the electron-electron Coulomb interplay would most likely modify the momentum transfer process. This may lead to the further modification of the overall angular distribution of the photoelectron given by $A(\mathbf{p})$. This would potentially affect the tilt parameters corresponding to the interference maxima. However, as shown in our previous discussion, it can be safely ruled out if we adopt the minima to calculate the tilt parameters.
On the other hand, the active electron remaining in the ion would change the deceleration of the photoelectron and thus might affect the interference term in Eq.~(\ref{eq:psim}). To estimate the impact, we may compare it to the case where an additional nucleus is present or not. 
For instance, the Coulomb attraction on the electron escaping from the first core at the beginning of the ionization for $R=1.4$ a.u.~is almost twice stronger than that for $R=10$ a.u. Yet, in both cases the deviation of the tilt parameters between the theoretical model and the numerical simulations (for the minima) is rather small, as shown in Fig.~\ref{fig:alpha}. Therefore, by qualitatively analyzing the electron-electron Coulomb interaction, we expect that the multielectron effect would have hardly impact on the locations of the interference minima.

\section{Conclusion and outlook}
In summary, we have shown how two kinds of nondipole effects modify the double-slit interference in single-photon ionization of H$_2^+$ when the xuv pulse travels in different directions. The single-atom nondipole effect leads to a tilted envelope of the overall interference pattern, whereas the interacting delay between the pulse and two nuclear centers further modifies the locations of the interference maxima and minima. Our results show that the locations of the interference minima are hardly changed by the single-atom nondipole effect. Further analysis indicates that the multielectron dynamics is likely to have insignificant effect on the interference minima as well. By adopting the two neighboring minima of the zeroth-order interference maxima to calculate the tilt parameters, we found good agreement between theoretical model and numerical simulations regarding the time delay between the pulse interacting with two centers of the molecule. 

Yet, the present work, as well as the previous theoretical study having included the multielectron dynamics \cite{z1}, fails to reproduce the experimental results in \cite{zept2}. 
In the framework of Schr\"odinger equation, the laser field is treated classically and we can investigate how the given laser field induces the electronic dynamics. However, how would the electronic motion impact the imposing photon during the {interaction}? From another perspective, there is electron cloud between two nuclei of a molecule, so is the phase velocity of the external field still the speed of $c$ inside the molecule? The answers to these questions are unknown yet. Further study is in progress as these questions might be the key to the gap between the measurement and the simple theoretical model regarding the zeptosecond time delay.

\section*{Acknowledgments}
Liu thanks Barth and Renziehausen for interesting discussion. This work is supported by the National Natural
Science Foundation of China (12174133, 92150106, 12021004). The computing work in this paper is supported by the public computing service platform provided by Network and Computing Center of HUST.

\end{document}